\pdfoutput=1
\RequirePackage{fixltx2e}

\documentclass[aps,floatfix,reprint,groupedaddress,showpacs,showkeys,amsmath,amssymb,]{revtex4-1}

\usepackage{amsfonts}%
\usepackage[ngerman,english]{babel}
\usepackage[utf8]{inputenc}
\usepackage{graphicx}
\usepackage{dcolumn}
\usepackage{bm}

\begin{document}


\title{Mott Variable Range Hopping and Weak Antilocalization Effect in Heteroepitaxial Na\textsubscript{2}IrO\textsubscript{3} Thin Films}


\author{Marcus Jenderka}
	\email[Corresponding Author: ]{marcus.jenderka@physik.uni-leipzig.de}
\author{José Barzola-Quiquia}
\author{Zhipeng Zhang}
\author{Heiko Frenzel}
\author{Marius Grundmann}
\author{Michael Lorenz}
	
\affiliation{%
Institute for Experimental Physics II\\
University of Leipzig\\
Linnéstraße 5, D-04103 Leipzig (Germany)
}%


\date{\today}

\begin{abstract}
Iridate thin films are a prerequisite for any application utilizing their cooperative effects resulting from the interplay of stron spin-orbit coupling and electronic correlations. Here, heteroepitaxial $\mathrm{Na_{2}IrO_{3}}$ thin films with excellent (001) out-of-plane crystalline orientation and well defined in-plane epitaxial relationship are presented on various oxide substrates. Resistivity is dominated by a three-dimensional variable range hopping mechanism in a large temperature range between 300 K and 40 K. Optical experiments show the onset of a small optical gap $E_{\mathrm{go}}$ $\approx$ 200 meV and a splitting of the Ir 5d-$t_{\mathrm{2g}}$ manifold. Positive magnetoresistance below 3 T and 25 K shows signatures of a weak antilocalization effect. This effect can be associated with surface states in a topological insulator and hence supports proposals for a topological insulator phase present in $\mathrm{Na_{2}IrO_{3}}$.
\end{abstract}


\maketitle


Transition metal oxides containing 5d iridium ions allow for the observation of novel cooperative effects resulting from an interplay between strong spin-orbit coupling and electronic correlations. These iridates are promising candidates for high-$T_{C}$ superconductors \cite{Wang2011b}, spin liquids \cite{Nakatsuji2006,Okamoto2007,Lawler2008}, a novel $J_{eff}$ = 1/2 Mott insulating ground state \cite{Kim2008,Kim2009} and for topological insulators \cite{Shitade2009,Pesin2010,Yang2010,Kim2012}.

A rather recently studied iridate is the Mott insulating layered compound Na\textsubscript{2}IrO\textsubscript{3} where edge-sharing IrO\textsubscript{6} octahedra form a honeycomb lattice within each Na\textsubscript{2}IrO\textsubscript{3} layer \cite{Singh2010}. Theoretical studies of magnetic interactions in model Hamiltonians of A\textsubscript{2}BO\textsubscript{3}-type compounds \cite{Jackeli2009,Chaloupka2010} suggest spin liquid behavior in Na\textsubscript{2}IrO\textsubscript{3}. On the other hand, tight-binding model analyses and first-principles band structure calculations \cite{Shitade2009,Wang2011,Kim2012}, as well as density-matrix renormalization group calculations \cite{Jiang2011} suggest Na\textsubscript{2}IrO\textsubscript{3} as a possible topological insulator. Both states of matter, however, promise application in fault-tolerant quantum computation \cite{Kitaev2003,Collins2006,Nayak2008}.

Experimental efforts on Na\textsubscript{2}IrO\textsubscript{3} were so far limited to powder and single-crystalline samples \cite{Singh2010,Choi2012,Ye2012,Liu2011,Comin2012}. Initially, from X-ray diffraction experiments a monoclinic $C2/c$ unit cell for Na\textsubscript{2}IrO\textsubscript{3} was suggested \cite{Singh2010}. More recent experiments however are more consistent with a $C2/m$ unit cell \cite{Choi2012,Ye2012}. Latter experiments also confirm the presence of trigonal distortions of the IrO\textsubscript{6} octahedra and that structural disorder, i.e. stacking faults and Na/Ir site mixings, is common. The compound furthermore exhibits frustrated antiferromagnetic order below $T_{\mathrm{N}}$ = 15 K with moments ordered collinearly in a zig-zag pattern \cite{Choi2012,Ye2012,Liu2011}. Furthermore, single-crystalline Na\textsubscript{2}IrO\textsubscript{3} has a small band gap \cite{Comin2012}. Its temperature dependent in-plane dc electrical resistivity follows an insulating $\rho\propto\exp[(T_{\mathrm{0}}/T)^{1/4}]$ behavior between 100 and 300 K \cite{Singh2010}. Such a $T$ dependence is usually associated with three-dimensional Mott variable range hopping of localized carriers.

In this communication, we report on the pulsed laser deposition (PLD) of heteroepitaxial Na\textsubscript{2}IrO\textsubscript{3} thin films on (001) YAlO\textsubscript{3}, a-sapphire, and c-sapphire. Deposition of Na\textsubscript{2}IrO\textsubscript{3} thin films ultimately is a step towards future device applications of this material. During PLD, we use growth temperatures of $T$ $\approx$ 550$^\circ$C and vary the oxygen partial pressure $p_{\mathrm{O2}}$ between 0.1 and 3.0 $\times$ 10$^{-4}$ mbar. Our heteroepitaxial films exhibit a clear epitaxial relation and an excellent (001) out-of-plane orientation. In magnetoresistance measurements we observe the weak antilocalization effect at 25 K and lower. Weak antilocalization can be associated with the surface states in a topological insulator \cite{Fu2007} and has previously been observed in established topological insulators such as Bi\textsubscript{2}Se\textsubscript{3} \cite{Chen2010,Chen2011,Kim2011,Taskin2012} and Bi\textsubscript{2}Te\textsubscript{3} \cite{He2011}.

Thin films were grown by PLD on 5 $\times$ 5 mm$^{2}$ and 10 $\times$ 10 mm$^{2}$ a-plane (11.0) sapphire, c-plane (001) sapphire and YAlO\textsubscript{3} (001) at temperatures and oxygen partial pressures ranging from about 550$^\circ$C to 650$^\circ$C and from 0.6 mbar to 3.0 $\times$ 10$^{-4}$ mbar, respectively. PLD was done with a 248 nm KrF excimer laser (Coherent Lambda Physik LPX 305i) at a laser fluence of 2 J$\,$cm$^{-2}$. The polycrystalline Na\textsubscript{2}IrO\textsubscript{3} target was prepared by a solid-state synthesis. According to Ref. \cite{Singh2010}, we homogenized Na\textsubscript{2}CO\textsubscript{3} (5N purity, alfa Aesar) and IrO\textsubscript{2} (85.45\% Ir content, ChemPUR) oxide powders in an atomic ratio of 1.05:1. The mixture was then calcined in a closed alumina crucible for 24 h at 750$^\circ$C in air. Afterwards, it was reground and pressed to obtain a cylindrical one-inch diameter target that in turn was sintered for 48 h at 900$^\circ$C. Higher sintering temperatures were not possible, since IrO\textsubscript{2} tends to sublimate above 1,000$^\circ$C. The deposition procedure involved a nucleation layer grown with 300 laser pulses at 1 Hz, followed by another 30,000 pulses at 15 Hz. After deposition, the samples were annealed in-situ at $p_{\mathrm{O2}}$ = 800 mbar. First ellipsometry measurements could only give a rough estimate of film thickness between 400 and 800 nm which, however, is still consistent with PLD growth rates for other oxide thin films \cite{Lorenz2013}.

For structural analysis we employed a Philips X’Pert X-ray diffractometer equipped with a Bragg-Brentano powder goniometer using divergent/focusing slit optics and Ni-filtered Cu K$_{\mathrm{\alpha}}$ radiation. Surface morphology was investigated via a Park System XE-150 atomic force microscope in dynamic non-contact mode. Temperature dependent dc electrical resistivity was measured in van-der-Pauw geometry using a Keithley 220 current source with a current range from 1 nA to 1 A. Transverse magnetoresistance was measured with a high resolution AC bridge (LR700 from Linear Research) in a commercial cryostat in the temperature range between 5 and 100 K with magnetic fields up to 8 T in both van-der-Pauw and four point geometry with sputtered gold contacts. Optical transmission and Fourier transform infrared spectroscopy were measured at ambient conditions and room temperature employing a Perkin Elmer Lambda 40 UV/VIS and a BRUKER IFS 66v/S Fourier spectrometer, respectively. Optical conductivity was obtained from current-voltage characteristics measured using an Agilent 4156C Precision Semiconductor Parameter Analyzer under illumination from a xenon arc lamp.

\begin{figure}
\includegraphics[width=0.75\columnwidth]{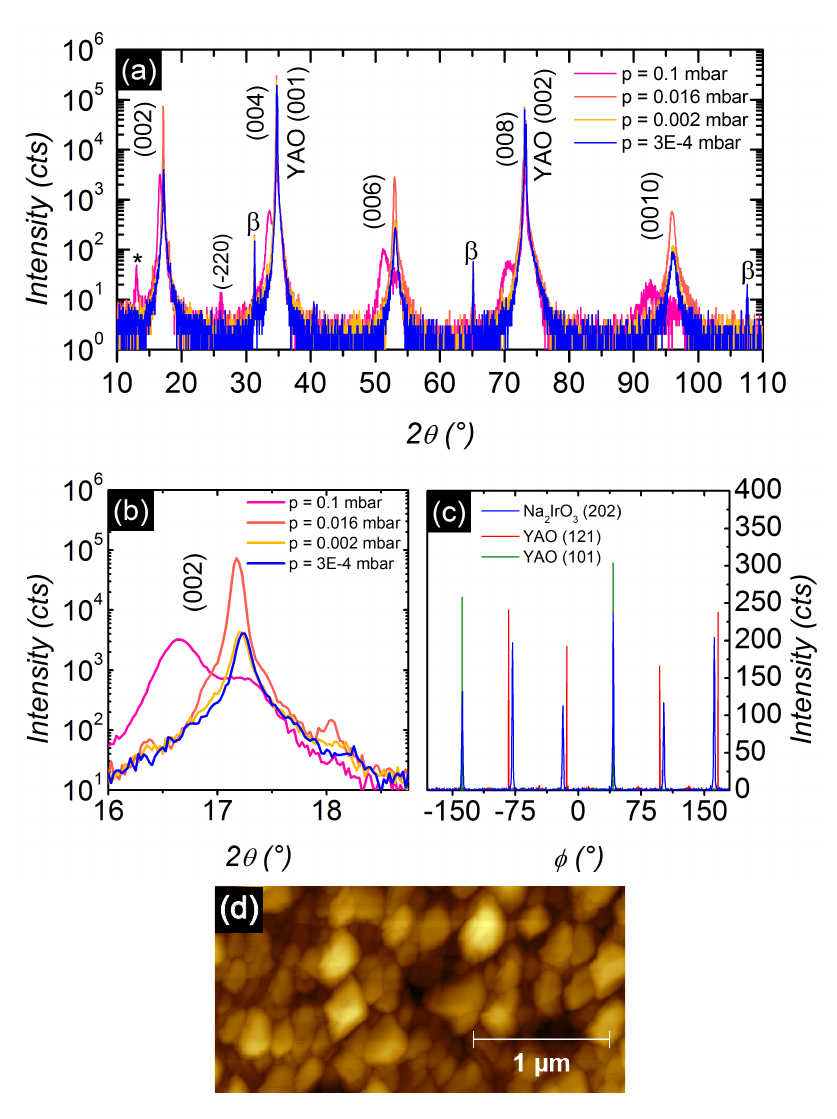}%
\caption{\label{fig:XRD}(Color online) PLD grown (001) oriented Na\textsubscript{2}IrO\textsubscript{3} films on YAlO\textsubscript{3} (001) (YAO). (a) XRD 2$\Theta$-$\omega$-scans of films grown at $p_{\mathrm{O2}}$ as indicated. (b) Zoom-in on the (002) reflections seen in (a). (c) Typical XRD $\phi$-scans of asymmetric Na\textsubscript{2}IrO\textsubscript{3} (202), YAO (101) and (121) reflections indicating the presence of six rotation domains ($p_{\mathrm{O2}}$ = 3.0 $\times$ 10$^{-4}$ mbar, $T$ $\approx$ 600$^\circ$C). (d) Non-contact AFM topographic image of a typical Na\textsubscript{2}IrO\textsubscript{3} film.}
\end{figure}

Structure and surface morphology of the as-grown thin films are investigated by X-ray diffraction (XRD) and atomic force microscopy (AFM). Typical XRD patterns of a series of Na\textsubscript{2}IrO\textsubscript{3} thin films grown on YAlO\textsubscript{3} (YAO) (001) at $T$ $\approx$ 550$^\circ$C and at oxygen partial pressures $p_{\mathrm{O2}}$ as indicated are displayed in Fig. \ref{fig:XRD}(a). The patterns are indexed according to the JCPDS diffraction database pattern 00-026-1376 for Na\textsubscript{2}IrO\textsubscript{3} in the monoclinic $C2/c$ unit cell. The patterns show very pronounced symmetric peaks related to the (001) planes of the Na\textsubscript{2}IrO\textsubscript{3} phase confirming its excellent out-of-plane preferential orientation. Two minor additional peaks with intensities below 10$^{2}$ counts can be related either to the (-220) orientation of the same phase or possibly to the (110) orientation of the closely related Na\textsubscript{4}Ir\textsubscript{3}O\textsubscript{8} phase (denoted as $*$ in Fig. \ref{fig:XRD}(a)). Decreasing $p_{\mathrm{O2}}$, we furthermore observe a clear tunability of the out-of-plane lattice parameter $c$ from 10.813 $\mathrm{\mathring{A}}$ to 10.435 $\mathrm{\mathring{A}}$ as can be seen in Fig. \ref{fig:XRD}(b). A variation in growth temperature has a comparably small effect on $c$ (see \cite{Jenderka2012} and supplemental material). In fact, the XRD results shown here also apply to our thin films grown on a-sapphire snd c-sapphire substrates. To illustrate the in-plane epitaxial relationship, Fig. \ref{fig:XRD}(c) shows $\phi$-scans of the asymmetric Na\textsubscript{2}IrO\textsubscript{3} (202), and the YAO (101) and (121) reflections. From the mismatch between rotational symmetry $C_{\mathrm{n}}$ and $C_{\mathrm{m}}$ of film and substrate, which is $C_{\mathrm{1}}$ for Na\textsubscript{2}IrO\textsubscript{3} and $C_{\mathrm{2}}$ for YAO, one expects to observe two rotational domains \cite{Grundmann2010}, i.e. two reflections in a $\phi$-scan of the Na\textsubscript{2}IrO\textsubscript{3} (202) reflection. Instead, we observe six reflections and assume that this increased number of rotational domains is either due to a mixed terminated surface of the YAO substrate containing half unit cell step heights \cite{Kleibeuker2010} or other nearly fulfilled additional symmetries of the substrate surface \cite{Grundmann2011}. The latter can be estimated for the YAO (121) and (101) reflections, as shown in Fig. \ref{fig:XRD}(c). A non-contact AFM scan of a thin film grown on YAO(001) is shown in Fig. \ref{fig:XRD}(d). Although XRD confirms the nearly perfect out-of-plane and in-plane epitaxy of our films, the 1.5 $\times$ 3 m$^{\mathrm{2}}$ topographic image reveals a granular surface structure with an RMS roughness at the given growth conditions of 13 nm. Grain sizes range in average from 100 to 200 nm, which is typical for oxide films.

\begin{figure}
\includegraphics[width=\columnwidth]{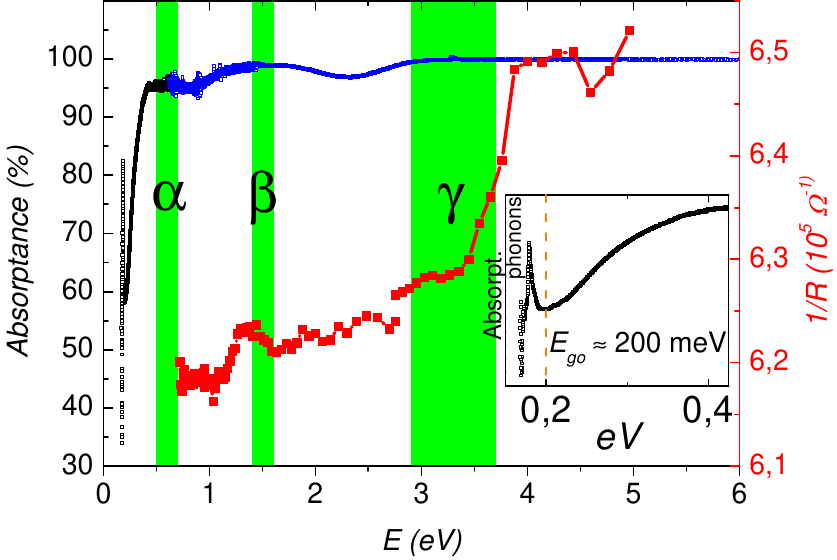}%
\caption{\label{fig:optics}(Color online) Optical absorptance measured with FTIR (black) and UV/VIS spectrometer (blue, both left scale) and optical conductivity under illumination with a Xe lamp (red, right scale) of Na\textsubscript{2}IrO\textsubscript{3} thin films at ambient conditions. Full measured spectral range from 0 to 6 eV shows three absorption peaks $\alpha$, $\beta$, $\gamma$ at 0.6, 1.5 and above 3 eV, respectively. The inset shows the FTIR data below 0.5 eV indicating a small optical gap of $E_{\mathrm{go}}$ $\approx$ 200 meV}
\end{figure}

A combination of Fourier transform infrared spectroscopy (FTIR), optical transmission and optical conductivity measurements on two samples, one grown on a-sapphire (FTIR), the other on c-sapphire (optical transmission and conductivity), is shown in Fig. \ref{fig:optics}. An energy range from 0 to 6 eV is covered by the measurements. In the FTIR data (black), we observe an absorption edge starting at $E_{\mathrm{go}}$ $\approx$ 200 meV indicating a small optical gap compatible with the recent finding \cite{Comin2012} of a 340 meV gap in Na\textsubscript{2}IrO\textsubscript{3} single crystals. We further observe indications of three distinct absorption peaks $\alpha$, $\beta$, $\gamma$ at 0.6, 1.5 and above about 3 eV, respectively. Based on recent angle-resolved photoelectron spectroscopy data of Na\textsubscript{2}IrO\textsubscript{3} single crystals \cite{Comin2012}, we assign features $\alpha$ and $\beta$ to intraband transitions within the split Ir $5d$-$t_{\mathrm{2g}}$ manifold, while feature $\gamma$ indicates interband transitions from Ir $5d$-$t_{\mathrm{2g}}$ into O $2p$ states. Similar transitions and small insulating gaps were previously observed in related materials Sr\textsubscript{2}IrO\textsubscript{4} \cite{Kim2008,Moon2008} and Ir\textsubscript{2}O\textsubscript{4} \cite{Kuriyama2010}. The splitting of the $t_{\mathrm{2g}}$ bands was proposed previously \cite{Comin2012,Jin2009,Bhattacharjee2012} due to an interplay between spin-orbit coupling, electronic correlations and possible trigonal distortions of the IrO\textsubscript{6} octahedra. Trigonal distortions in turn are deemed neccessary to facilitate the experimentally observed \cite{Choi2012,Ye2012} zig-zag type antiferromagnetic order.

\begin{figure*}
\includegraphics[width=\textwidth]{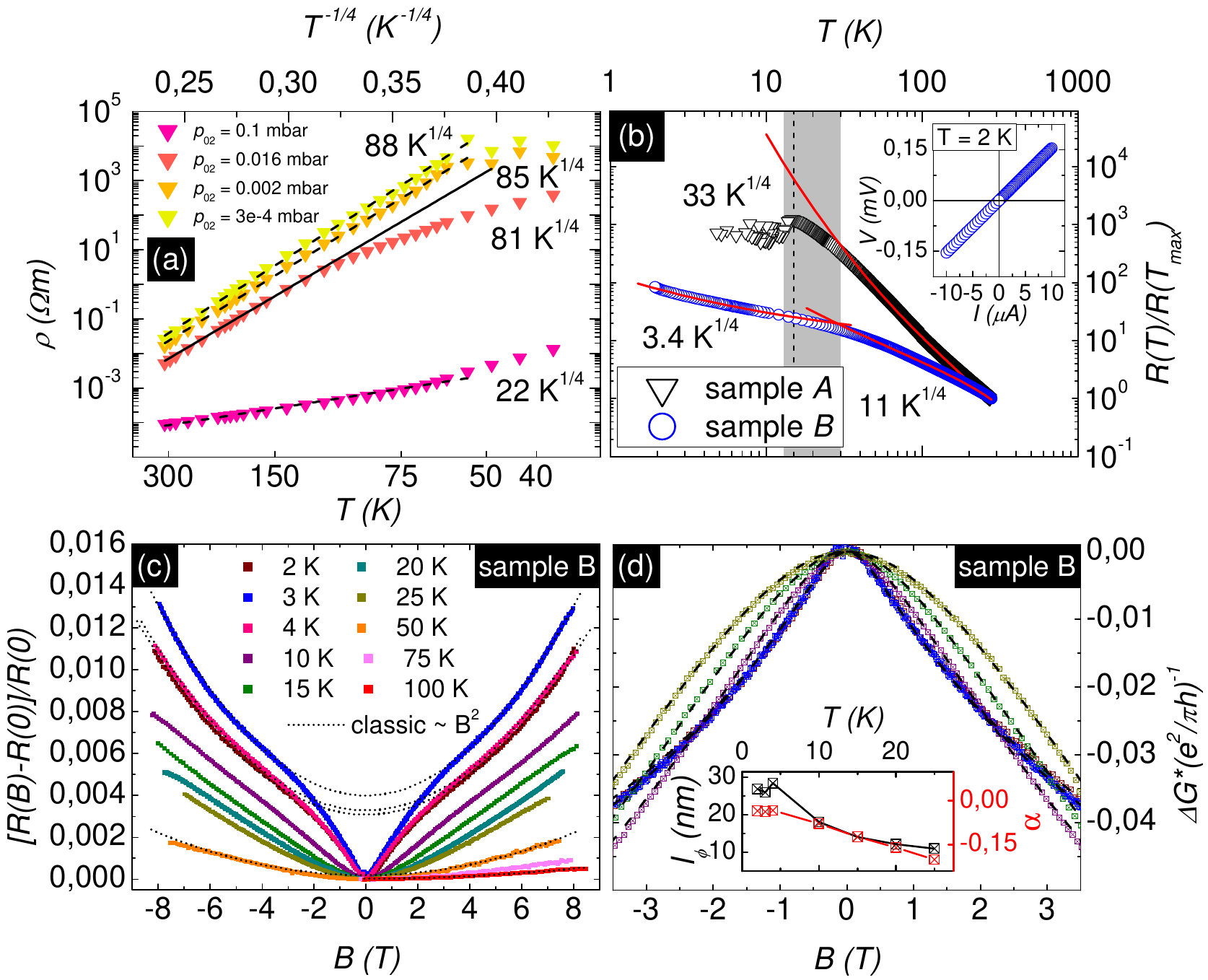}%
\caption{\label{fig:transport}(Color online) Transport properties. (a) Temperature-dependent resistivity of Na\textsubscript{2}IrO\textsubscript{3} thin films on YAO(001). Data are plotted in $\log\rho$ versus $T^{-1/4}$. Straight line fits with the respective slopes $T_{\mathrm{0}}^{1/4}$ indicate Mott-VRH conductivity mechanism. (b) Resistance $R$ versus $T$ plotted in log-log scale measured from 280 K to 2 K of two samples $A$, $B$ (see text). Sample $A$ shows an anomaly around 15 K (dashed line) related to reported antiferromagnetic ordering temperatures $T_{\mathrm{N}}$ of Na\textsubscript{2}IrO\textsubscript{3} single crystals. Red lines are fits according to Mott-VRH model. Inset: current-voltage characteristic of sample $B$ at 2 K. (c,d) Magnetoresistance of sample $B$. WAL behavior in a Na\textsubscript{2}IrO\textsubscript{3} film on c-sapphire. (c) Normalized out-of-plane magnetoresistance $\Delta R/R(0)$ between 2 and 100 K. In large fields $B$ $\geq$ 4 T, experimental data follows a classical parabolic law (dotted lines). (d) The same data expressed as normalized magnetoconductivity $\Delta G\cdot(e^{2}/\pi h)^{-1}$ is fit by the HLN equation for $T$ $\leq$ 25 K and magnetic fields $B$ $\leq$ 3 T (dashed lines, see text). The inset illustrates the temperature dependence of the fit parameters $\alpha$ and $l_{\mathrm{\phi}}$.}
\end{figure*}

For our thin films, we investigate the influence of $p_{\mathrm{O2}}$ on temperature-dependent electrical resistivity $\rho$. Figure \ref{fig:transport}(a) shows the electrical resistivity $\log\rho$ versus $T^{-1/4}$ between 30 K and 300 K for a series of Na\textsubscript{2}IrO\textsubscript{3} thin films grown on YAO(001) at $T$ $\approx$ 550$^\circ$C. For all $p_{\mathrm{O2}}$, the films show semiconducting behavior. For thin films grown at $p_{\mathrm{O2}}$ = 0.1 mbar, resistivity at 300 K is around 2.4 $\times$ 10$^{-4}$ $\mathrm{\Omega}$m. It monotonously increases to 2.8 $\times$ 10$^{-2}$ $\mathrm{\Omega}$m for $p_{\mathrm{O2}}$ = 3 $\times$ 10$^{-4}$ mbar, indicating the tunability of resistivity (see Table \ref{tab:table1}). At $p_{\mathrm{O2}}$ = 0.002 mbar and lower, resistivity exceeds our measuring range up to 6 M$\mathrm{\Omega}$ at low temperatures. The temperature dependence of resistivity between 40 and 300 K can be described using a three-dimensional Mott variable range hopping (VRH) model \cite{Mott1969}, for which
\begin{equation}
\rho=\rho_{\mathrm{0}}\exp[(T_{\mathrm{0}}/T)^{1/4}].
\label{eq:MottVRH}
\end{equation}
This is indicated by the straight line fits in Fig. \ref{fig:transport}(a). Thus, we observe VRH similarly as previously observed in Na\textsubscript{2}IrO\textsubscript{3} single crystals \cite{Singh2010}. The localization temperature $T_{\mathrm{0}}$ of the VRH model is given by \cite{Shklovskii1984}
\begin{equation}
T_{\mathrm{0}}=21.2/k_{\mathrm{B}}a^{3}N(E_{\mathrm{F}}),
\label{eq:MottT0}
\end{equation}
where $k_{\mathrm{B}}$ is Boltzmann’s number, $a$ is the localization length, i.e. the decay radius of its wavefunction, and $N(E_{\mathrm{F}})$ is the density of states at the Fermi level \cite{Mott1969}. VRH is usually associated with the localization of carriers by disorder. Fitting the experimental resistivity results using eq. (1), we obtain $T_{\mathrm{0}}$ comparable with various other transition metal oxides \cite{Kastner1988,Bremholm2011,Colman2012,Yildiz2008,ADMA:ADMA201103087} (see Table \ref{tab:table1}). Using eq. (2) it is then possible to calculate $a$ or $N(E_{\mathrm{F}})$. However, knowledge of either $a$ or $N(E_{\mathrm{F}})$ is required. Here, we assume a constant $N(E_{\mathrm{F}})$ in the order of 10$^{28}$ eV$^{-1}$m$^{-3}$. We estimated $N(E_{\mathrm{F}})$ from heat capacity measurements performed on various iridates \cite{Bremholm2011,Zhao2009,Kini2006,Hinatsu2001,Carter1995}, where the coefficient $\eta$ was obtained from a fit of $C/T$ = $\eta$ + $\beta T$ at low temperatures. In these materials, the coefficient $\eta$ ranges from 0.5 to 10 mJ$\,$K$^{-2}\,$mol-Ir$^{-1}$ and is related to $N(E_{\mathrm{F}})$ via $\eta$ = $\pi^{2}k_{\mathrm{B}}^{2}V_{\mathrm{m}}N(E_{\mathrm{F}})/3$, with $V_{\mathrm{m}}$ being the molar volume of Ir. The localization lengths $a$ calculated from the fitted $T_{\mathrm{0}}$ and constant $N(E_{\mathrm{F}})$ using eq. (2) are in the order of 1 $\mathrm{\mathring{A}}$ and given in Table \ref{tab:table1}. Their magnitude is reasonable in comparison with the reported Ir-Ir and Ir-O bond distances in Na\textsubscript{2}IrO\textsubscript{3} of about 3 and 2 $\mathrm{\mathring{A}}$, respectively \cite{Choi2012,Ye2012}. According to our calculation we observe a correlation between $a$ and the resistivity, i.e. as $a$ decreases we observe a dramatic increase in resistivity $\rho$.
\begin{table}[b]
\caption{\label{tab:table1}%
Na\textsubscript{2}IrO\textsubscript{3} thin films grown on YAO(001) at $T$ $\approx$ 550$^{\circ}$C and oxygen partial pressures $p_{\mathrm{02}}$: room temperature resistivity $\rho$($T$ = 300 K), $T_{\mathrm{0}}$ is a fit parameter according to eq. (\ref{eq:MottVRH}). $a$ was calculated using eq. (\ref{eq:MottT0}) assuming $N(E_{\mathrm{F}})$ = 10$^{28}$ eV$^{-1}$m$^{-3}$.}
\begin{ruledtabular}
\begin{tabular}{lccc}
\textrm{$p_{\mathrm{O2}}$ (mbar)}&
\textrm{$\rho$($T$ = 300 K) ($\mathrm{\Omega}$m)}&
\textrm{$T_{\mathrm0}$ (K)}&
\textrm{$a$ ($\mathrm{\mathring{A}}$)}\\
\colrule
0.01 & 8.0 $\times$ 10$^{-5}$ & 2.3 $\times$ 10$^{5}$ & 4.72\\
0.016 & 5.2 $\times$ 10$^{-3}$ & 4.4 $\times$ 10$^{7}$ & 0.82\\
0.002 & 1.6 $\times$ 10$^{-2}$ & 5.2 $\times$ 10$^{7}$ & 0.78\\
3 $\times$ 10$^{-4}$ & 2.6 $\times$ 10$^{-2}$ & 6.0 $\times$ 10$^{7}$ & 0.74\\
\end{tabular}
\end{ruledtabular}
\end{table}

Figure \ref{fig:transport}(b) shows normalized resistance $R(T)/R(T_{\mathrm{max}})$ between 280 and 2 K of two samples $A$ and $B$ grown at 0.016 mbar and $T$ $\approx$ 550$^\circ$C on YAO(001) and c-sapphire, respectively. $R$ was measured by van-der-Pauw and four point method for samples $A$ and $B$, respectively. Sample $B$ was furthermore capped by a 110 nm thick SiN\textsubscript{x} layer and used for magnetoresistance measurements discussed in the next section. The current-voltage characteristic of sample $B$ at 2 K, shown in the inset of Fig. \ref{fig:transport}(b), is a straight line highlighting the excellent contact quality and ohmic behavior persisting down to 2 K. Quantitative differences between the resistance curves of samples $A$ and $B$ most likely arise from the different substrate used. For both samples, a 3D Mott-VRH model describes the data between 280 and 50 K, however, with different slopes $T_{\mathrm{0}}^{1/4}$ = 33 K$^{1/4}$ ($A$) and $T_{\mathrm{0}}^{1/4}$ = 11 K$^{1/4}$ ($B$). In addition, sample $A$ displays a sharp change in $R$ around 15 K which can be related to the formation or suppression of scattering processes which decrease the resistance for $T$ $<$ 15 K. A similar effect is also observed in sample $B$ below 25 K. It is known from some antiferromagnetic (AF) materials, that resistance diminishes below the Néel temperature $T_{\mathrm{N}}$ \cite{Fote1970,Jiang1994,Lorenzer2012}. For single-crystalline Na\textsubscript{2}IrO\textsubscript{3}, literature reports on the formation of an AF phase with a Néel temperature $T_{\mathrm{N}}$ between 13.4 and 18.1K \cite{Singh2010,Choi2012,Ye2012,Liu2011}. We suggest that the decrease of the resistance below 15 and 25 K in our investigated samples $A$ and $B$, respectively, is a consequence of the AF phase formation, possibly with a reduction in spin scattering. For sample $B$ we further observe that the low temperature resistance ($T$ $<$ 25 K) can also be well described by the Mott-VRH dependence, but with a much smaller slope of $T_{\mathrm{0}}^{1/4}$ = 3.4 K$^{1/4}$ compared to the high temperature range. With regard to the formation of AF long-range order and assuming the same $N(E_{\mathrm{F}})$ as before, this decrease of $T_{\mathrm{0}}$ can be related to an increase in the electron localization length $a$ from 12 to 56 $\mathrm{\mathring{A}}$. However, these values are one order of magnitude larger in contrast to the samples discussed above and suggest that a constant $N(E_{\mathrm{F}})$ for all samples is merely a simplifying assumption. As will be discussed below, we similarly observe a significant increase in the phase coherence length $l_{\mathrm{\phi}}$ extracted from magnetoresistance for $T$ $\leq$ 25 K. At present, it is not understood why Mott-VRH is observed within such an extended range of temperatures in both our epitaxial films and in single crystals \cite{Singh2010}. Theories on variable range hopping \cite{Mott1969,Shklovskii1984,Ambegaokar1971} require the localization of states within a narrow band near the Fermi level. In this spirit, the localization of states due to strong on-site Coulomb interactions of the Ir $5d$ orbitals and structural disorder induced by frequent stacking faults and interatomic site mixings \cite{Singh2010,Choi2012} would be consistent with variable range hopping behavior. In addition, the existence of narrow Ir $5d$-$t_{\mathrm{2g}}$ bands \cite{Comin2012} and a small insulating gap (cf. Ref. \cite{Comin2012} and $E_{\mathrm{go}}$ $\approx$ 200 meV in this work), would support variable range hopping.

Out-of-plane magnetoresistance (MR) measurements performed on two samples between 2 K and 100 K are shown in Fig. \ref{fig:transport}(c). The normalized magnetoresistance $R/R(0) = [(R(B) - R(0))/R(0)]$ is positive for all measured temperatures. In fields $B$ $>$ 3 T, MR scales well with $B^{2}$ which can be ascribed to the typical Lorentz contribution \cite{Ziman2007}. In fact, MR for $T$ $\geq$ 75 K follows this parabolic dependence for all measured fields. However, at low temperatures and at fields $B$ $<$ 3 T we observe a quick rise in MR. This behavior can be understood according to the weak antilocalization (WAL) effect. In light of recent proposals of topologically non-trivial phases present in Na\textsubscript{2}IrO\textsubscript{3} \cite{Shitade2009,Kim2012,Jiang2011,Wang2011}, the observation of WAL is very interesting, as it has been observed in thin films of established topological insulators Bi\textsubscript{2}Se\textsubscript{3} \cite{Chen2010,Chen2011,Kim2011,Taskin2012} and Bi\textsubscript{2}Te\textsubscript{3} \cite{He2011}. In a system with strong spin-orbit coupling, WAL originates from the destructive interference of coherently back-scattered conduction electrons due to spin rotations \cite{Bergmann1984}. WAL is also associated with topological surface states \cite{Fu2007}: Surface electrons acquire a Berry phase of $\pi$ leading to destructive quantum interference, as well. In both cases, conductance is enhanced due to the suppression of backscattering. In a magnetoresistance experiment, the magnetic field partially destroys the destructive quantum interference and leads to an unusual rise in resistance. Experimentally, we study the observed WAL effect by fitting the low-field magnetoconductivity (MC) for $T$ $\leq$ 25 K shown in Fig. \ref{fig:transport}(d) with the Hikami-Larkin-Nagaoka (HLN) equation for a 2D system in the limit of strong spin-orbit coupling \cite{Hikami1980}:
\begin{equation}
\Delta G(B)=-\alpha\frac{e^{2}}{\pi h}\left[\ln(\frac{\hbar}{4el_{\mathrm{\phi}}^{2}B})-\Psi(\frac{\hbar}{4el_{\mathrm{\phi}}^{2}B})\right],
\label{eq:HLN}
\end{equation}
where $\Psi$ is the digamma function and $l_{\mathrm{\phi}}$ denotes the phase coherence length. The prefactor $\alpha$ is expected equal to -1/2 for both a traditional 2D system with strong spin-orbit interactions \cite{Hikami1980} and \emph{one} surface of a topological insulator \cite{Chen2010}, i.e. for both the top and bottom surface of a topological insulator thin film without contribution from the bulk one expects $\alpha$ = -1. Our experimental data can be fit well by eq. (1) for fields $B$ $\leq$ 3 T and $T$ $\leq$ 25 K as the dashed lines in Fig. \ref{fig:transport}(d) illustrate. With temperature, the cusp continuously broadens. The inset shown in Fig. \ref{fig:transport}(d) displays the temperature dependence of the extracted fit parameters $\alpha$ and $l_{\mathrm{\phi}}$. Opposed to $\alpha$ = -1/2 corresponding to one surface electron channel, our fitted $\alpha$ decreases from -0.034 to -0.20 with increasing temperature. We also note the small values of the phase coherence length $l_{\mathrm{\phi}}$ decreasing from 27 nm at 2 K to 11 nm at 25 K. From our fit results we infer, that in our films WAL is reduced in comparison with experiments on established topological insulator thin films \cite{Chen2010,Chen2011,Kim2011,Taskin2012,He2011}, where film thickness is often below 50 nm, $\alpha$ ranges from -0.38 to about -1 and $l_{\mathrm{\phi}}$ ranges from 100 nm to 1,000 nm. In our films, the thickness is in the 100 nm regime, such that a dominating bulk contribution to transport properties is very likely. Deviations from $\alpha$ = -1/2 can be attributed to the scattering on magnetic impurities which can lead to a reduction of WAL bringing $\alpha$ closer to zero \cite{Hikami1980,He2011,Lukermann2012,Liu2012}. Such magnetic impurities could either be inherent to the bulk or caused by contamination during sample processing. Small $l_{\mathrm{\phi}}$ have been suggested as evidence for reduced screening and increased electron-electron interaction effects in a regime of low charge carrier densities \cite{Chen2010}. We argue that similar detrimental effects on the phase coherence length $l_{\mathrm{\phi}}$ can be caused by defects at the interface or increased impurity scattering due to high film thickness. It is also theorized, that bulk channels with opposite effect, i.e. weak localization causing negative MR, could partially compensate the WAL of the surface channels \cite{Lu2011}. At this point however, a quantitative discussion on the observed WAL is merely speculative. First thickness dependent transport measurements (not shown here) indicate, however, that conductivity below 40 K is nearly independent on film thickness thus hinting at a surface dominated conduction in this range of temperatures.

We have demonstrated that heteroepitaxial Na\textsubscript{2}IrO\textsubscript{3} thin films with very good out-of-plane crystalline orientation and defined in-plane epitaxial relationship can be grown by PLD on various oxide substrates. Resistivity is dominated by three-dimensional variable range hopping. Optical experiments indicate a small optical gap $E_{\mathrm{go}}$ $\approx$ 200 meV and a splitting of the Ir $5d$-$t_{\mathrm{2g}}$ manifold. For positive magnetoresistance below 3 T and 25 K we observed signatures of a weak antilocalization effect expected for topological insulators. The discovery of such an effect in Na\textsubscript{2}IrO\textsubscript{3} thin films is very intriguing as it supports propositions \cite{Shitade2009,Kim2012,Jiang2011,Wang2011} of a topologically non-trivial phase in this material. Na\textsubscript{2}IrO\textsubscript{3} hence remains a promising candidate to extend the range of topological insulator materials.

\begin{acknowledgments}
We thank the German DFG for financial support within the project LO790/5-1 "Oxide topological insulator thin films". The work of JBQ was supported by the
Collaborative Research Center SFB 762 "Functionality of Oxide Interfaces"
\end{acknowledgments}

\bibliography{library}

\end{document}